# Inquiring the Potential of Evoking Small-World Properties for Self-Organizing Communication Networks

Steffen Rothkugel and Matthias R. Brust
*University of Luxembourg*
*Faculty of Science, Technology and Communication (FSTC)*
*Luxembourg*
*{Steffen.Rothkugel, Matthias.Brust}@uni.lu*

Carlos H.C. Ribeiro
*Technological Institute of Aeronautics*
*Department of Computer Science*
*Brazil*
*carlos@comp.ita.br*

**Abstract**

*Mobile multi-hop ad hoc networks allow establishing local groups of communicating devices in a self-organizing way. However, in a global setting such networks fail to work properly due to network partitioning. Providing that devices are capable of communicating both locally—e.g. using Wi-Fi or Bluetooth—and additionally also with arbitrary remote devices—e.g. using GSM/UMTS links—the objective is to find efficient ways of inter-linking multiple network partitions. Tackling this problem of topology control, we focus on the class of small-world networks that obey two distinguishing characteristics: they have a strong local clustering while still retaining a small average distance between two nodes. This paper reports on results gained investigating the question if small-world properties are indicative for an efficient link management in multiple multi-hop ad hoc network partitions.*

## 1. Introduction

Inter-linking multiple multi-hop ad hoc network partitions allows overcoming the limitations inherent in pure ad hoc networks [1]. However, it is necessary to pay particular attention on the efficiency of those inter-links. Consequently, the dynamic and self-organizing augmentation of networks by additional links plays an important role. The costs of such augmentation should usually be made as small as possible.

Our approach is based on the hypothesis that small-world properties are promising indicators for a good set of rules to maximize inter-link efficiency. Before motivating deeply this approach a short introduction is given in the following section.

The subsequent section explains some background of injection networks [1] that are used to build our network model. In the subsequent section experiments that have been conducted to investigate qualitative and quantitative aspects of small-worlds in ad hoc networks are discussed. As one methodology we use empirical results from probabilistic analysis techniques employing random replacements of edges, similar to most previous studies of small-world networks [2]. For the analysis, a software tool called *ConnyLab* is used and is described and validated in detail in [3]. Initial discussions about how to tune the parameters, e.g. which nodes to inter-link, to maximize the benefit finalize this paper.

## 2. Small-Worlds

### 2.1 Introduction

Many classes of *regular* networks e.g. square grid networks or the restricted class of *circular networks* studied in [2] have a high *clustering coefficient*, i.e. nodes have many mutual neighbors, but large *characteristic path length*, i.e. large average shortest path length between pairs of nodes. The formal definition is:

**Definition 2.1 (Watts 1999)** *The local clustering coefficient $\gamma$ of one node $v$ with $k_v$ neighbors is*

$$\gamma_v = \frac{|E(\Gamma_v)|}{n(n-1)/2}, \quad (1)$$

*where $|E(\Gamma_v)|$ is the number of links in the neighborhood of $v$ and $n(n-1)/2$ is the number of all possible links. The clustering coefficient $\gamma$ is than the average local clustering coefficient for all nodes.*

**Definition 2.2 (Watts 1999)** *Calculate the shortest path length $\bar{d}_v$ connecting each node $v \in V(N)$ of a network $N$ to all other nodes, this is $d(v,j) \forall j \in V(N)$. Then define the characteristic path length $L$ as the median of all shortest paths.*

The opposite extreme to regular networks are *random* networks, which have a small characteristic path length but exhibit very little clustering. Networks between these two extremes can be constructed by starting with a regular network and randomly moving one endpoint of each edge with probability *p*. Regular networks correspond to *p* = 0 and random networks are obtained by setting *p* = 1. Watts and Strogatz [4] discovered that the characteristic path length decreases rapidly as *p* increases, but the clustering coefficient decreases slowly. It was shown that the initial topology and the construction are not important to arrive at the essential characteristic of a small-world network [2]. Thus, for any given network, it is possible to determine whether or not it is a small-world graph without knowing anything of its construction.

A small-world network must show a specific correlation between characteristic path length and clustering coefficient (small-world properties). In that case, we may classify the network to obey small-world properties by comparing it with the initial topology where *p* = 0 (cf. construction above). However, there are disadvantages following this approach. In particular, there is the need for finding an appropriate construction, but also producing the whole family of possible topologies. According to an alternative, but equivalent approach [4], a small-world graph is a graph with *n* vertices and an average degree *k* that exhibits characteristic path length of $L \approx L_{Random}(n,k)$, but clustering coefficient of $\gamma >> \gamma_{Random}(n,k) \approx k/n$.

Observe that our network model deals with *transitional networks* (a combination between *spatial* and *relational networks*), but the work of Watts is generally based on pure relational networks only. In fact, it was shown that in some cases small-world properties also exhibits in spatial networks [5], but if this happens in transitional networks is still an open question, in particular, for the general case.

Some researchers which combine ad hoc networks with the idea of small-worlds disregard the difference between relational, spatial, and transitional networks [6, 7] or don't care about the required correlation between characteristic path length and clustering coefficient for small-worlds [7].

## 2.2 Small-Worlds in Self-Organizing Networks

After presenting the basics of small-worlds in the last section, we now argue why small-worlds can be beneficial if applied in e.g. ad hoc networks.

As pointed out in Herrmann et al. [8], "mobile ad hoc networks provide a communication environment that is characterized by dynamic changes in the topology and in the availability of resources … [that] make it difficult to transfer well-known mechanisms for the organization of distributed systems to the ad-hoc networking domain". Herrmann et al. conclude that applications are able to cope with these effects by using self-organization mechanisms.

Without going deeper into the various definitions that exist, we may use a more intuitive concept of self-organization that already has been used by [9-11]: *entropy*. Entropy is a measure of the disorder of a system. Systems tend to go from a state of order (low entropy) to a state of maximum disorder (high entropy). Self-organizing principles aim at keeping the system's entropy as low as possible. Gershenson et al. [10] point out that the entropy is strongly dependent on the evaluation perspective. Thus, "an ant colony will not be self-organizing if we describe only the *global* behavior of the colony (e.g. as an *element* of an ecosystem), or if we only list the behaviors of individual ants. We have to remember that the description of self-organization is partially, but strongly, dependent on the observer."

Additionally, we consider that self-organizing systems own intrinsic characteristics like complexity, redundancy, autonomy etc. [12].

In the domain of mobile communication, technologies like for instance Bluetooth and Wi-Fi are able to create communication links within the transmission range at no charge. Additionally, e.g. GMS- or UMTS-adapters can be employed to establish supplementary costly communication links between two arbitrary devices. The aim of adding and removing these supplementary links is to arrive at an "efficient" network topology obeying small-world properties, i.e. high local clustering together with small characteristic path length. Because of the possibility to dynamically control such links, the network topology basically can be biased as needed. Thus, this approach is able to compensate the disadvantages caused by mobility, in fact realizing a topology control.

The dynamic augmentation of an ad hoc network by carefully adding special links is the anchor point for self-organizing principles to maintain low entropy. Of course, obsolete links might also be removed.

We motivate to measure the entropy from the "perspective" of small-world networks, since typical data-flow patterns in communication networks show a large amount of clustering with a small number of "long distance" communications that need to be completed quickly, what in fact is the characteristic of a small-world graph.

To show its generic nature, we illustrate this data-flow pattern also in a mobile ad hoc networking-based application. Let us consider a typical m-learning scenario as proposed in [13]. In this application, lecture material, slides, annotations, and particularly also tasks are initially distributed to the students. A common behavior is to form groups and resolve tasks locally, but since not all tasks can be tackled within the own group, usually other groups are consulted. In other words, for successfully running this application we need to design a communication network that owns the two distinguishing characteristics of small-world networks: strong local clustering (group) and small average distance between two nodes (asking someone outside the own group). Hence, the communication pattern corresponds so far with our approach.

Furthermore, characteristic path length and clustering coefficient also have a concrete significance related to multi-hop ad hoc networks. A small path length corresponds in fewer hops what is definitely important for routing mechanisms and the overall communication performance of the entire network. Local clustering matches with redundancy requirements and reflects the power of information dissemination of nodes. These described circumstances strongly motivate the use of small-worlds.

Obviously, the described ad-hoc communication network environment also shows practical factors as nodes' instability, low and different transmission range, low bandwidth, missing Bluetooth or GSM-/UMTS-adapters, low storage capabilities and limited battery power, and more. However, even with these limitations in mind, we consider them as being of minor importance for our investigations at this moment. Nevertheless, special links must in fact be used carefully to limit overall costs or the energy consumption.

## 3. Injection Networks

Due to several problems that are inherent to mobile multi-hop ad hoc networks, some approaches focus on hybrid wireless networks where a fixed infrastructure supports a higher connectivity and avoid network partitioning [14]. Often, however, such an *infrastructured* hybrid wireless network isn't feasible, because of economical as well as implementing issues. Alternatively, an *infrastructureless* setting is of interest where problems of restricted geographical regions are avoided.

Helmy [15] focuses on long-range links. The objective is to reduce the number of queries during the search for a given target node. Another approach introduces base stations to increase connectivity in ad hoc networks [14], thus realizing global reachability. Watts [2] introduces a spatially defined link, called *global edge*, with length-scaling properties to include spatial models in his investigations. Some approaches suggest as network model a standard ad hoc network, but extending it by two different transmission ranges [7, 12], e.g. small distance Bluetooth links together with higher distance Wi-Fi links.

Our initial motivation for the current investigation is based on the assumption that technologies like Bluetooth and Wi-Fi can be employed to create ad hoc communication links within the transmission range at no charge.

Additional cellular network links such as GSM/UMTS might be employed by appropriately equipped devices to establish supplementary communication links between two arbitrary devices [1]. These links, however, will induce costs. Furthermore, we follow the transparency paradigm, that linking in a mobile multi-hop ad hoc network should be managed without explicit human interaction, i.e. self-organized.

In summary, different approaches exist to augment ad hoc networks with additional links for different reasons: increasing connectivity, augment network capacity, boost bandwidth between particular devices, and also to inter-link multiple ad hoc network partitions (cf. [16]). Creating a generic entity for this entire links, we introduce the term *bypass link*.

**Definition 3.1 (Watts)** *The spatial neighborhood $\Gamma_{tr}^{s}(v)$ of a node v is the set of nodes within transmission range tr of v.*

**Definition 3.2** *A bypass link is a link $(u,v)$ between nodes u and v with $u \notin \Gamma_{tr}^{s}(v)$.*

That is, a bypass link is a link that connects two nodes that are not in the same spatial neighborhood. Please note that elements of $\Gamma_{tr}^{s}(v)$ do not necessarily have to be connected to *v* in real settings.

Practically, a bypass link can be realized over a cellular network as well as access points. Nevertheless, in our considerations a bypass link is counted as a single hop, albeit this finally ignores the real topology behind that bypass link.

Because of dynamically controlling such bypass links, the network topology basically can be biased as needed, resulting in a topology control.

The injection communication paradigm as described in [1] is based on establishing bypass links between carefully selected devices. Herrmann et al. [8] called these dedicated communication points *hub nodes*. Depending on the overall objective, the selection of such devices can be driven by different factors, like for instance to obtain a high clustering coefficient, or devices with certain attributes. Such attributes might for instance be information available on a particular device, or also services offered. These dedicated devices used for establishing bypass links are called *injection points* in [1]. For self-organizing communication networks based on bypass links and injections points as described before we use the term *injection networks*.

**Definition 3.3** *If a bypass link $(u,v)$ between nodes u and v exists, both nodes u and v are called injection points.*

Injection points serve two different purposes: a point where information dissemination starts and where services are being placed (*service placement* [8]).

In the first case, the injection point is of essential importance at the moment of receiving information and passing this information to the neighborhood.

The injection point might represent a bottleneck, depending on the amount of data passing through. Injection points become particularly attractive when offering a service. Information dissemination can be seen as such a service that is usable by devices in the injection point's surroundings.

Different criteria for determining the injection point can be of importance. Considerations include available power, technical equipment, load balancing issues, the time a device is expected to remain available, and also the information currently available on a device. Supposing that for instance the device is highly clustered and thus one of the central members of a group, epidemic behavior for information spreading will take effect faster. Therefore, the current environment and the device's relationship to its neighbors are important.

As mentioned already, in this paper we focus on structural criteria that are given by the small-world properties: characteristic path length and clustering coefficient. The injection point has to react accordingly to the changing network structure. Based on its predictions and on the current network structure, the hosting device has probably to be changed for re-fulfilling the selection criteria.

## 4. Experiments

In this section, three indicative experiments are conducted that investigate the potential of small-worlds in injection networks. The first experiment gives an idea of the communication efficiency of a pure ad hoc network topology compared to a regular grid. In the second experiment, the potential gain for a network setting with a single bypass link is investigated. In the first and second experiment qualitative aspects are investigated. In the third experiment the quantitative aspect of bypass links is explored, analyzing the potential when using multiple injection points. In each experiment, we deal with small network partitions of fifty nodes each, which is a sound partition size for the application scenarios under investigation.

### 4.1. Ad hoc Net Construction

In order to arrange characteristic ad hoc network topologies in a correspondent graph family, a spatial network growth construction is implemented. As illustrated in Fig. 1 the construction starts with a finite regular square lattice for probability $p = 0$ and tends to result in a fully random topology for $p = 1$. The detailed construction is as follows:

1. Initialize $p = 0$
2. Create a regular square lattice of $n$ nodes
3. For each node $n_i$ relocate the node with probability $p$. If the relocation of node $n_i$ would result in a disconnected network, go to the next node $n_{i+i}$
4. Connect $n_i$ to all nodes within its transmission range $tr$. Node $n_i$ must be connected to at least one neighboring node, otherwise search a new location and repeat step 4
5. Increase $p$ until $p \geq 1$, repeating step 2

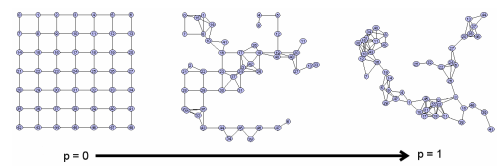

**Fig. 1.** Schema of the spatial construction algorithm. For $p = 0$ the original grid is unchanged; for $p = 1$, all edges are rewired and all nodes are replaced randomly showing ad hoc network topology characteristics; for $0 < p < 1$, graphs combining elements of order and randomness are generated

Note the importance of the constraint in step 4 to guarantee fully connected graphs, because $L$ is only defined for fully connected graphs. Observe also that the number of new neighbors might vary for each node.

An example of the construction is illustrated in Fig. 1. This construction arranges ad hoc network topologies in a family of networks that though allows comparing its properties between them.

Nevertheless there are other constructions that would fulfill this criterion. Observe that it still has to be shown how far this result corresponds to existing probability density functions as used in geometric random graphs theory, occupancy theory, and continuum percolation. Finally, $L$ and $\gamma$ are measured (cf. Fig. 2).

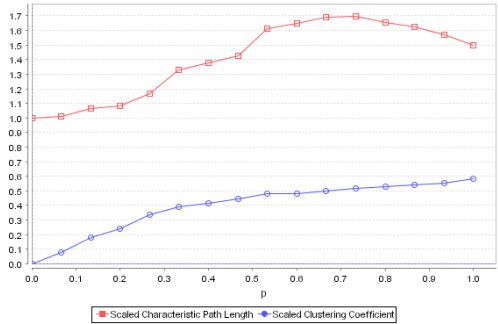

**Fig. 2.** Normalized $L$ and $\gamma$, $n = 100$, averaged over 20 runs

**Results.** Looking at Fig. 1, it becomes apparent that $p = 1$ results in topologies representing ad hoc networks as they can appear in shopping malls, marketplaces, or train stations. Interestingly, in Fig. 2 it seems that maintaining a regular or semi-regular topology results in a better network performance with respect to $L$. In compensation, $\gamma$ is getting higher by increasing $p$. Concluding, these measurements show that an ad hoc network topology is a quite inefficient communication network in terms of hops, but more efficient in terms of clustering than the regular grid serving as a starting point.

### 4.2. Using one Bypass Link

Taking two ($p = 1$)-family members from the construction above, an important question is what happens with $L$ and $\gamma$ when introducing a single bypass link to inter-connect them? Please note that now the network model obeys transitional characteristics as described in Section 2.

The network partitions are composed of fifty nodes each. The transmission range of each device is $tr = 50$ units. Devices that are within the transmission range are connected.

Additionally, the two network partitions are connected by exactly one bypass link, as illustrated in Fig. 3. In each step the bypass link is reassigned in such a way that it connects both network partitions by two arbitrarily chosen injection points. Please note that the normalization value related to L is approximated.

**Results.** The resulting chart shows $L$ varying between 0.62 and 0.97 indicating a potential difference of around 33%. High values for $L$ indicate two badly chosen injections points, while low values for $L$ designate two favorable injections points. Aside from that, the 33%-potential difference also could have been approximately be confirmed by conducting experiments for n = 3, 5, 10, 20, 25, 50, and 100. The value for $\gamma$ doesn't vary much, staying between 0.64 and 0.66. This is because a single bypass link only influences the local clustering coefficient of exactly two nodes.

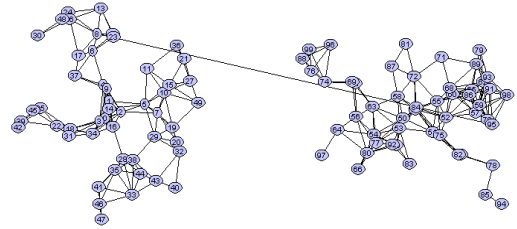

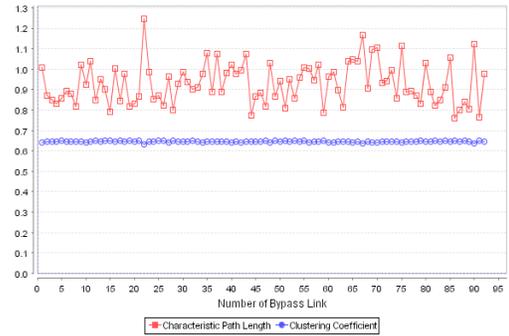

**Fig. 3.** Network transition (above): Two ad hoc networks are inter-linked by one bypass link using two arbitrarily chosen injection points in each step. Measurements of $L$ and $\gamma$ in the chart below

After 25 runs an average $k$ of 7.512 is determined. As defined in the introduction the correspondent value of the clustering coefficient of a random network is $\gamma_{Random}(n,k) \approx k/n$ that is with $n = 100$ and $k = 7.512$ exactly 0.07512. Consequently, the condition $\gamma \approx 0.65 \gg \gamma_{Random}(100, 7.512) \approx 0.07512$ holds true meaning that $\gamma$ fulfills the small-world criteria for relational graphs even in our transitional network model. We also need to investigate the second condition $L \approx L_{Random}(n,k)$ that is important for small-worlds as well. Calculations show that $L_{Random}(100, 7.512) \approx 0.29$ which is rather different from $L \approx 0.62$. To take a discussion in advance, we take this back to (a) a small $n$, since the approximation works with larger $n$, (b) the spatial factor, and (c) that the experiment includes a very small fraction of bypass links. Concluding, the introduction of multiple injection points appears necessary.

### 4.3 Multiple Injection Points

Due to mobility and fast topology changes, a sparse ad hoc network or multiple network partitions causes problems in efficiently spreading information or providing services. However, employing several simultaneous injection points can be supposed to result in a more reliable setting to fulfill the injection network as well as the small-world criteria. The previous experiment showed that the bypass link—as relational factor—can influence $L$ considerably, but has not much influence on $\gamma$. Hence, instead of reassigning a single bypass link, additional ones are added. Is it possible this way to further improve $L$ fulfilling the small-world condition but without degrading $\gamma$ too much?

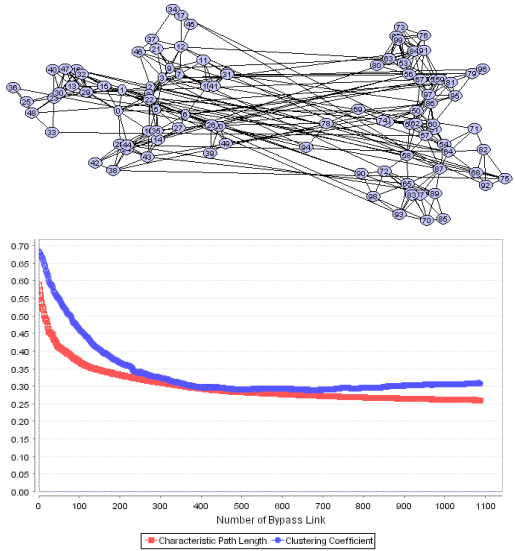

**Fig. 4.** Network transition (above): Two ad hoc networks are inter-linked by an increasing number of bypass links. Measured data for evaluation in a chart (below)

**Results.** Concerning the characteristic path length, as illustrated in Fig. 4, the more bypass links are added, the lower $L$ will be. For a small numbers of bypass links $b$ the value of L will decrease fast. Starting at $b = 30$ the process slows down, indicating that a reasonable number of bypass links for this network configuration has been reached. Further augmenting this number will not increase benefits significantly.

Concerning the clustering coefficient, $\gamma$ decreases almost proportional with $L$ until $b = 300$. This is due to the definition of $\gamma$. As long as the number of bypass links is comparatively small, adding a single bypass link increases the number of possible links significantly, thus lowering $\gamma$. Interestingly, after passing $b = 300$, $\gamma$ starts increasing again slowly. This happens because very much bypass links have been established already, and hence the number of possible additional links is increasing slower than the number of existing links.

Even more important, $L$ decreases around $b = 300$ to 0.29 while $\gamma \approx 0.32 \gg \gamma \approx 0.07512$ still is true. Hence, it seems possible evoking small-world properties in ad hoc networks using a certain number of bypass links. Please note that by adding bypass links the value of $k$ naturally increases slowly, so that this issue has to be observed further. But the results give reason to interpret these values as valid approximation, further motivating additional investigations, also analytical ones. On the other hand, injection points are chosen uniformly in a random way, i.e. both well and badly chosen injection points. The second experiment, however, indicates that a favorable value for $L$ can be obtained with significantly fewer bypass links when choosing injection points carefully, while almost preserving the value of $k$.

## 5. Future Work: Evoking Small-World Properties

We evaluated the second experiment in a way that a higher value is achieved when well-chosen injection points are connected by a bypass link. However, the experiments conducted up to now do not give information about the characteristics of such promising injection points. Nevertheless, by repeating the experiments a couple of times, we got an idea of such characteristics. To capture this, heuristics are being developed, aiming at describing good and bad (local) situations axiomatically.

Additionally, we showed that the small-world properties seem to be evocable, but based upon global information. In practice, however, the aim is to avoid a central infrastructure collecting and integrating the necessary information. Deciding locally is the preferred approach, relying on information from as few devices in the neighborhood as possible. The general feasibility as well as the quality to be expected is subject of current research.

## 6. Conclusions

Small-world properties can be found in the World Wide Web, social networks, and even in nervous systems. In the case of ad hoc networks, it is possible through the use of bypass links—connections that inter-link two arbitrary nodes disregarding transmission range and network partitioning—to evoke small-world features. Three experiments dealing with qualitative and quantitative aspects of bypass links and injection points are discussed in this paper.

We discovered a potential difference of about 33%, meaning that through well-chosen injection points the number of average hops can be reduced to around 67%. This potential that has to be explored further, keeping in mind the numerous packets transmitted in a network. Especially in mobile multi-hop ad hoc networks where devices act as routers, spending much energy in forwarding messages, the overall power consumption can be decreased by carefully chosen injection points. Certainly, based on the promising results discussed throughout this paper, it is worth to design an appropriate model for analytical investigations and to validate our empirical results.

Consequently, the next steps are finding approaches for fostering the evocation of small-world features in a way that the values converge to an optimum. Node mobility introduces another level of complexity. Due to dynamic and often unpredictable topology changes, injection points also need to be reassigned over time.

Injection-based communication networks (*injection networks*) exhibit a natural form of self-organization that manifests itself in their internal structure, revealing a noteworthy improvement in the communication efficiency. Their inherent potential still needs to be unleashed. In this paper we have shown that there is a considerable capacity in evoking small-world properties in self-organizing communication networks.